# Geopolitical Tensions and Financial Networks: Strategic Shifts Toward Alternatives

Antonis Ballis[*]

## Abstract

Global financial systems are undergoing strategic shifts as geopolitical tensions reshape international trade and payments. The United States (US)–China trade war, sanctions regimes, and rising concerns over the weaponization of financial infrastructures like SWIFT have led countries to seek alternative networks, including China's CIPS and emerging cross-border CBDCs. This letter presents a dynamic theoretical framework where sanction risks, investment choices, and network effects drive payment system migration. Empirical evidence from Russia, Saudi Arabia, India, and Argentina supports the model. Policy implications point toward increasing financial fragmentation, with critical roles for international institutions to mitigate systemic risks. The future of finance may be less global and more regionally fragmented, influenced heavily by political considerations.

---

[*] Aston Business School, Aston University, Birmingham, UK. Email: a.ballis@aston.ac.uk. (Corresponding Author)

# 1. Introduction: Trade Wars, Sanctions, and the Financial System

Since 2018, the intensifying US-China trade war has exposed vulnerabilities at the core of the global financial architecture. Sanctions have evolved from narrow diplomatic tools to strategic levers capable of weaponizing financial infrastructures like SWIFT (Cipriani et al., 2023). Earlier precedents, notably Iran's exclusion from SWIFT in 2012, foreshadowed the broader use of financial access as a geopolitical instrument (Efing et al., 2023). Russia's partial expulsion from SWIFT following the Ukraine invasion underscored the reality: control over financial messaging networks confers decisive strategic advantages (Farrell and Newman, 2019).

In response, major economies have accelerated efforts to build alternatives. China expanded its Cross-Border Interbank Payment System (CIPS); emerging economies like India and Argentina established local currency settlement mechanisms; central banks launched cross-border CBDC pilots. Simultaneously, trade war dynamics evolved beyond tariffs into financial, technological, and resource-based containment strategies, with US export controls on semiconductors and China's curbs on critical minerals like gallium (2023) emblematic of a broader decoupling trend.

These shifts reflect a growing imperative among states to pursue financial sovereignty: reducing exposure to adversarial control over payment channels and reserve currencies. This letter develops a theoretical framework linking sanctions risk, investment behaviour, and migration across payment systems under network effects. Empirical evidence from Russia, Saudi Arabia, India, and Argentina shows that rising geopolitical tensions systematically reshape financial behaviour, potentially fragmenting the global payments ecosystem into politically aligned blocs.

# 2. Theoretical Framework: Sanctions Risk, Investment, and Payment System Choice

In an increasingly polarised geopolitical environment, countries, firms, and financial institutions face strategic decisions about their dependence on dominant financial infrastructures. To capture this dynamic, we develop a simple but powerful theoretical framework linking sanction risks, investment choices, and payment system switching under network effects.

Consider an agent choosing between two systems: System $S$ (such as SWIFT), which is highly efficient but vulnerable to sanctions, and System $A$ (such as CIPS or CBDC-based platforms), which offers resilience at the cost of lower initial efficiency. If the agent remains with System $S$, it faces a baseline sanction probability $p_0$. However, the agent may choose to invest effort $e \geq 0$ to reduce this probability. The effectiveness of investment is linear, captured by

$$p(e) = p_0 - \alpha e,$$

where $\alpha > 0$ represents the degree to which investment reduces sanction risk. The cost of investing is convex, given by

$$C(e) = \frac{1}{2} k e^2,$$

where $k > 0$ reflects the cost severity.

The agent's expected utility from remaining with System $S$, after choosing an optimal level of investment $e$, can be written as:

$$EU_S(e) = \big(1 - p(e)\big)(1 + \epsilon + \theta N_S) + p(e)(1 + \theta N_S - L) - C(e)$$

where $\epsilon > 0$ captures the efficiency advantage of $S$ over $A$, $L > 0$ represents the loss incurred if sanctions are triggered, $\theta > 0$ measures the strength of network externalities, and $N_S$ denotes the share of agents still using System $S$.

Alternatively, switching to System $A$ yields a stable payoff:

$$EU_A = 1 + \theta N_A$$

where $N_A$ is the share of agents using the alternative system. Notably, $N_A$ is initially small, meaning that early switchers suffer from reduced network benefits, making coordination between agents critical.

The decision facing the agent is a two-stage problem. At $t = 0$, the agent must decide whether to invest in risk mitigation and remain with $S$, or immediately switch to $A$. At $t = 1$, sanctions are realised probabilistically, and based on aggregate behaviour, the network shares $N_S$ and $N_A$ evolve. Early movers pay a network cost by joining a thinner system, but mass migration to $A$ over time increases $N_A$, enhancing its attractiveness to others.

Coordination frictions arise because of these network effects. Even if the geopolitical risk $p_0$ increases, individual agents may hesitate to switch if others are not yet moving. This inertia implies the existence of a critical mass phenomenon: once enough agents migrate, the switching cost diminishes, inducing rapid further adoption of the alternative system.

Formally, agents will prefer switching to $A$ if:

$$EU_A > \max_e E\, U_S(e)$$

This defines a critical sanction probability threshold $p^*$ beyond which agents collectively abandon the incumbent system. The threshold $p^*$ depends on several parameters: it declines with the sanction loss $L$, increases with the efficiency advantage $\epsilon$ of System $S$, rises with the strength of network effects $\theta$, and depends inversely on the evolving shares $N_S$ and $N_A$. Crucially, $p^*$ is not static: it changes dynamically as early switching alters the network composition.

Thus, even small shocks (i.e. a new sanctions package, a major country defecting to an alternative system, a breakthrough in CBDC technology reducing $\epsilon$) can move the system past a tipping point, unleashing rapid systemic migration. This theoretical structure not only predicts gradual erosion of incumbent systems under rising geopolitical stress but also suggests that shifts, when they occur, may appear sudden and nonlinear due to underlying coordination frictions.

While the framework above captures the key trade-offs driving agents' decisions, the analysis can be extended by formally endogenising network externalities and examining the resulting equilibrium dynamics. This deeper treatment illustrates how coordination frictions and non-linear feedback mechanisms can give rise to tipping points in system migration. For completeness, this extended version of the model is provided in Appendix.

## 3. Evidence: Shifting Trade and Payment Patterns

The theory that geopolitical tensions push financial actors toward alternative systems is not merely hypothetical; it is unfolding in real time across several major economies. Nowhere has this dynamic been more vivid than in Russia's rapid financial pivot after the imposition of SWIFT sanctions in 2022. Within just two years, the share of Russian exports invoiced in Chinese yuan exploded from under 3% to over 30%. Russian banks, cut off from Western clearing channels, turned swiftly to China's CIPS network, and yuan liquidity surged (Drott et al., 2024) as Moscow recalibrated its external trade architecture. CIPS itself recorded a sharp uptick in activity, with total settlement volumes growing by more than 20% in 2022 alone, reaching nearly 97 trillion RMB.

This shift was not isolated. Saudi Arabia, long seen as a bedrock of the petrodollar system, entered into active discussions with China in 2023 about accepting RMB for oil sales. While formal agreements have been tentative, the mere willingness to consider pricing oil in yuan signals a profound strategic reconsideration. A similar logic guided India's decision to operationalise a rupee–ruble trade settlement mechanism with Russia. Confronted with the prospect of sanctions spillovers and dollar dependence, Indian policymakers moved quickly to insulate key bilateral trade flows from external disruptions.

Argentina's experience offers another striking example. Facing a crippling shortage of US dollars, Buenos Aires activated a swap line with China to finance Belt and Road imports directly in RMB (Keerati, 2023). By mid-2024, the yuan had become a significant medium of settlement for Argentine foreign trade, with over $10 billion in transactions bypassing the traditional dollar-based system.

Even in regions traditionally aligned with the West, cracks are visible. In 2023, the United Arab Emirates and India agreed to settle their burgeoning trade relationship in dirhams and rupees, a move emblematic of a broader global trend: local currency arrangements designed to circumvent the risks associated with relying solely on dollar-centric networks.

Across these diverse cases, a consistent pattern emerges. Actors facing elevated geopolitical risks, whether through direct sanctions, secondary exposure, or broader strategic rivalry, increasingly seek alternatives. Early adopters initially bore higher switching costs due to limited network size. But as participation grew, the benefits of joining alternative systems compounded, gradually reducing the friction for new entrants. Migration toward alternative financial networks is rarely immediate or linear, but once critical mass is achieved, shifts can accelerate with remarkable speed.

***Figure 1***

This shift is further illustrated by the rising global usage of the Chinese RMB for cross-border payments. As shown in Figure 1, the RMB's share in SWIFT transactions nearly doubled between 2018 and 2023, reflecting growing adoption of alternative settlement currencies in response to geopolitical frictions. While SWIFT remains the primary messaging infrastructure, the underlying choice of invoicing currency has begun to diversify, laying the groundwork for potential future migration to independent systems such as CIPS.

## 4. Implications: Fragmentation, CBDCs, and Global Finance

The cumulative effect of these movements is the slow but perceptible fragmentation of the global financial system. Where once a single unified network facilitated most cross-border transactions, largely anchored around SWIFT and dollar liquidity, we are witnessing the emergence of parallel ecosystems. One centred on Western institutions, another forming around China-led initiatives like CIPS, CBDCs, and localised bilateral settlement frameworks.

This splintering comes at a cost. Higher transaction costs, duplicated compliance regimes, and reduced liquidity depth are inevitable consequences (Hafner-Burton et al., 2020). Financial actors who once enjoyed seamless access to global payment rails must now navigate an increasingly complex and politically sensitive landscape. Emerging economies, in particular, find themselves balancing between networks, facing higher operational uncertainty as they hedge their exposures.

Central Bank Digital Currencies (CBDCs) introduce another dimension of strategic complexity. China's digital yuan, already piloted across borders in projects like mBridge with Hong Kong, Thailand, and the United Arab Emirates (UAE), exemplifies the ambition to build payment systems decoupled from SWIFT's gravitational pull. India's digital rupee and Saudi Arabia's digital rial pilots serve a similar purpose: creating resilient, politically autonomous settlement channels that can function even under conditions of financial fragmentation.

Meanwhile, the data reveals deeper structural shifts underway. By late 2023, the dollar's share of global foreign exchange reserves fell to its lowest level in decades, dropping to just 58.4%. This trend is clearly reflected in the official reserve compositions reported by the IMF, as shown in Figure 2. The erosion of dollar dominance, once seen as a distant prospect, is now measurable, albeit gradual. Functional de-dollarization is emerging not through dramatic upheaval but through a steady accumulation of alternative practices, local currency settlements, CBDC pilots, and RMB-denominated trade agreements (Chen et al., 2024).

***Figure 2***

Financial infrastructures, once judged primarily on efficiency and liquidity, are now increasingly assessed through the lens of sovereignty, resilience, and political alignment. Payment systems have become instruments of statecraft. The result is a world moving not toward deeper financial integration, but toward a patchwork of interoperable, politically aligned systems, each shaped by strategic necessity as much as by technological innovation.

Managing this transition will be one of the most significant challenges of international finance in the coming decades. Interoperability projects, like SWIFT's CBDC initiatives, may soften the edges of fragmentation, but they cannot erase the underlying strategic drivers. The future of global finance will be defined not by universal convergence, but by the ability of different systems to coexist, sometimes awkwardly, sometimes competitively, in a world where trust, sovereignty, and resilience have become as important as speed and efficiency.

5. Conclusion

Geopolitical tensions have accelerated the transformation of global financial networks. Our theoretical model shows how sanctions risk, network frictions, and dynamic investment decisions jointly drive strategic migration toward alternative systems. Empirical evidence from Russia, Saudi Arabia, India, Argentina, and others reveals that theory and reality are converging. As risks rise, actors reassess the trade-offs between efficiency and political resilience, often choosing diversification even at economic cost. The future of global finance is unlikely to be monolithic. It will be shaped by fragmented, politically aligned infrastructures where efficiency, sovereignty, and trust interact in complex ways. Managing the transition, through interoperability initiatives, neutral platforms, and strengthened governance, will be one of the defining challenges of international finance in the coming decades.

**Appendix. Endogenising Network Externalities and Equilibrium Dynamics**

While the baseline model treats network effects as a function of current system shares, the structure can be enriched by explicitly modelling how these externalities evolve endogenously and shape agent behaviour in equilibrium.

Let $s_{A(t)}$ and $s_{B(t)} = 1 - s_{A(t)}$ denote the fraction of agents using System $A$ and $B$ at time $t$, respectively. Instead of assuming linear or fixed network benefits, we define them as increasing functions of usage shares:

$$\theta(s) = \alpha s^{\gamma} \text{ with } \gamma > 1$$

This formulation captures increasing returns to adoption: early adopters of System B face low network benefits, but as more agents migrate, the utility from joining rises non-linearly.

The agent's expected utility from remaining in $A$, including mitigation effort $z$, becomes:

$$U_A = \epsilon - p(z)L + \theta(s_A) - C(z)$$

The utility from switching to $B$ depends only on the share of other users:

$$U_B = \theta(s_B)$$

An agent is indifferent between staying and switching when:

$$\epsilon - p(z^*)L + \theta(1 - s_B^*) - C(z^*) = \theta(s_B^*)$$

This defines a critical share $s_B^*$: if enough others switch, the incentive flips.

To model evolution, consider a replicator dynamic:

$$\dot{s_B} = s_B(1 - s_B)[U_B - U_A]$$

This yields multiple equilibria: when $s_B$ is low, agents prefer to remain with system $A$ (status quo); at the tipping point where $s_B = s_B^*$, the system is in an unstable equilibrium; and when $s_B$ is high, agents converge toward System $B$.

Small increases in sanction risk $p$, improvements in $B$'s technology (lowering the efficiency gap $\epsilon$), or a decline in mitigation effectiveness $\delta$, can nudge the system past this tipping point, triggering a rapid migration. This endogenous treatment better reflects observed non-linear switching behaviour and complements the baseline model's policy intuition by showing how network evolution, not just individual utility, drives systemic change.

**Figures**

**Figure 1:** RMB Share in SWIFT Global Payments (2018–2023).

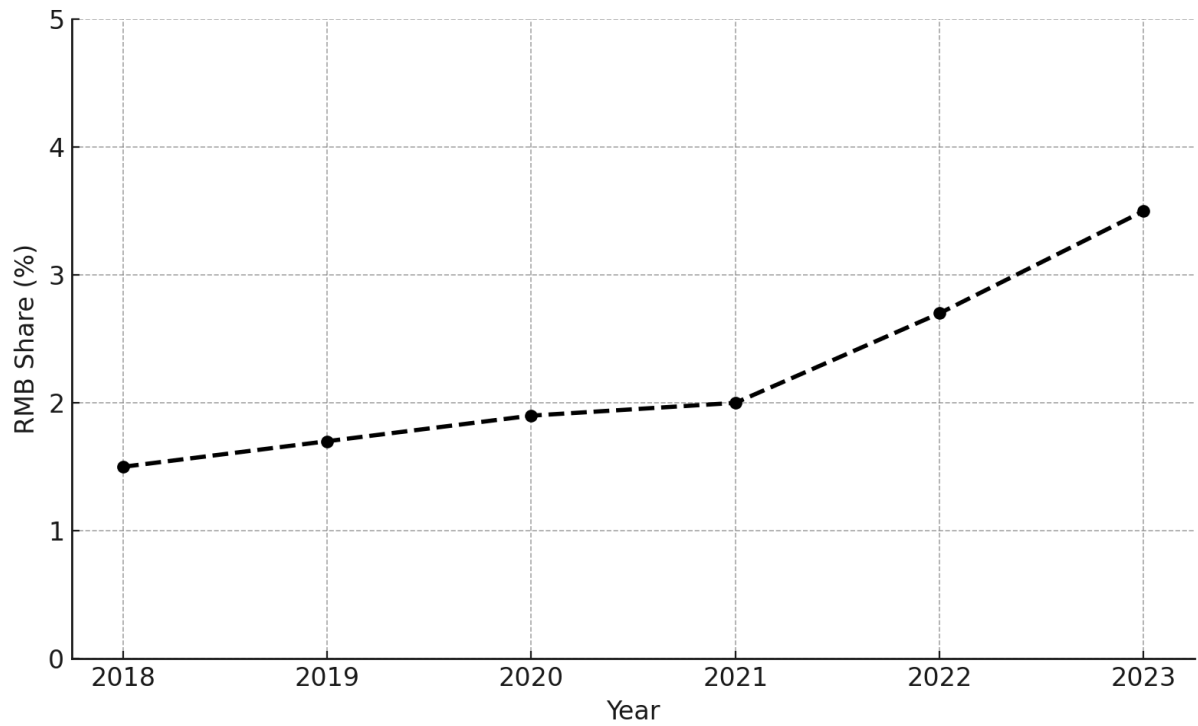

Source: SWIFT RMB Tracker reports, compiled from annual summaries.

**Figure 2:** USD Share of Global Foreign Exchange Reserves (2000–2023).

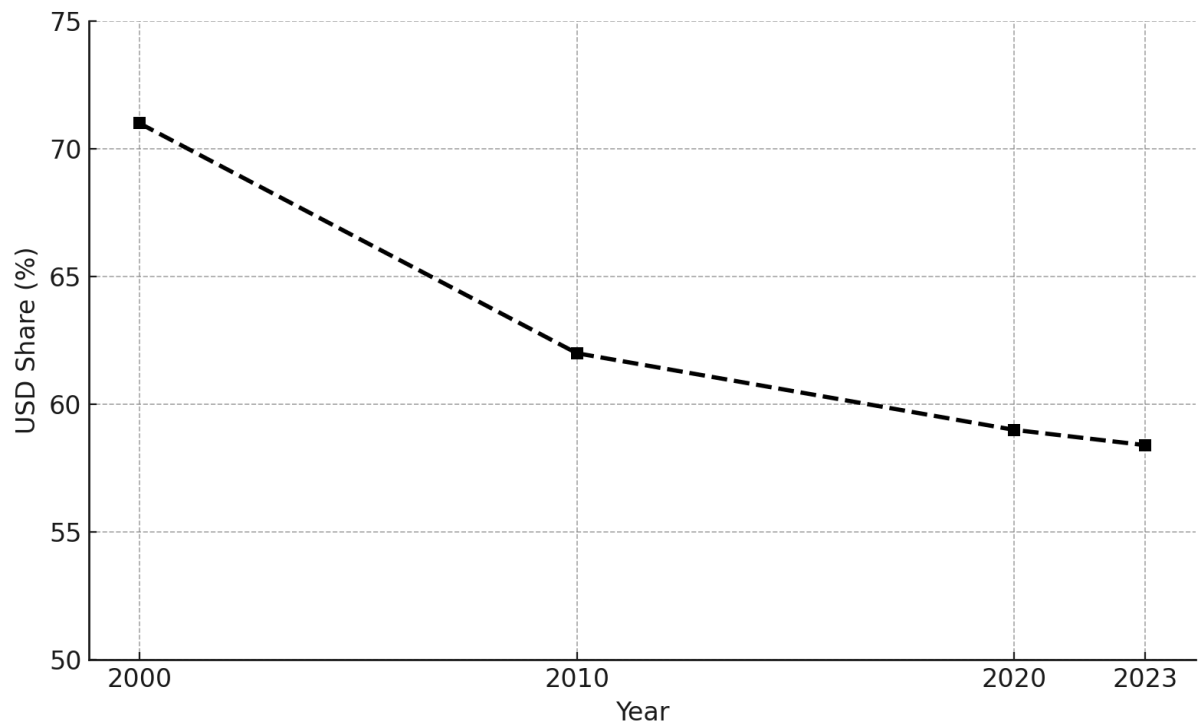

Source: IMF COFER Database (2023). Data compiled from IMF quarterly reserve reports.